# Rule-based Construction of Matching Processes

[Technical Report]

Eric Peukert
SAP Research Center Dresden
01187 Dresden, Germany
eric.peukert@sap.com

Julian Eberius
Dresden University of Technology
Dresden, Germany
julian.eberius@tu-dresden.de

Erhard Rahm
University of Leipzig
Leipzig, Germany
rahm@informatik.unileipzig.de

## ABSTRACT
Mapping complex metadata structures is crucial in a number of domains such as data integration, ontology alignment or model management. To speed up that process automatic matching systems were developed to compute mapping suggestions that can be corrected by a user. However, constructing and tuning match strategies still requires a high manual effort by matching experts as well as correct mappings to evaluate generated mappings. We therefore propose a self-configuring schema matching system that is able to automatically adapt to the given mapping problem at hand. Our approach is based on analyzing the input schemas as well as intermediate matching results. A variety of matching rules use the analysis results to automatically construct and adapt an underlying matching process for a given match task. We comprehensively evaluate our approach on different mapping problems from the schema, ontology and model management domains. The evaluation shows that our system is able to robustly return good quality mappings across different mapping problems and domains.

## Categories and Subject Descriptors
D.2.12 [**Interoperability**]: Data mapping

## General Terms
Algorithms

## 1. INTRODUCTION
Finding mappings between complex metadata structures is a critical task in a number of domains such as data integration, ontology alignment or model transformation. We call this task schema matching, but it can also be named ontology alignment [18] or metamodel matching [9]. In order to speed up that task, semi-automatic schema matching systems were developed. These systems rely on matching algorithms, so called matchers, to compute a mapping suggestion that can be corrected by a user. A multitude of schema matching systems and matching algorithms were proposed (see [22], [21] or [1] for overviews). Except for some domain-specific matchers, the algorithms used in the different systems are often similar, e.g. they consider the linguistic and structural similarity of schema elements or the similarity of instance data. Many systems are constructed for a single schema type or domain and may even be tuned for specific benchmarks such as the OAEI Benchmark [8].

Constructing and tuning a schema matching system is a complex, manual and time-consuming task. It often requires a lot of matching experience and expert knowledge as well as given perfect mappings. Tuning is either done manually in code, or is sometimes supported by special user interfaces for configuration as proposed in [3] or [1]. Schema matching publications typically report the maximally achieved quality of automatically computed mapping suggestions using some specially tuned parameter configuration.

However, such an approach cannot be adopted for applying a schema matching system in practice onto fully unknown matching problems. Moreover, users often do not have schema matching experience, so that they rely on default match configurations, i.e. a predetermined selection of matchers and combination of their match results. Unfortunately, these default configurations are often not robust enough to cope with largely differing matching problems of diverse domains. Hence, there is a need for adaptive and robust matching systems that return good mappings across different matching problems without manual tuning.

There have already been some attempts to make parts of a matching system [11][14][20][16] more adaptive and self-tuning. For instance, the ontology alignment system RiMOM [14] computes two properties of the input schemas to later select or unselect a structural and a name-based matcher. However, these adaptations are fixed in the code and seem to be optimizations tailored to the OAEI benchmarks.

In this paper, we propose a more comprehensive approach for a fully self-configuring schema matching system that can automatically construct and adapt a matching process for a given mapping problem. Specifically, we make the following contributions:

- We introduce so called features that are computed from the input schemas as well as from intermediate mapping results. Among others, we introduce the so called monogamy feature that allows predicting the quality of a mapping result without having a correct mapping.
- Based on the features, we introduce several *matching rules* that represent expert knowledge on how to define or adapt a schema matching process. In particular, we introduce a rule for automatically finding parameters for the selection operator that selects the final match correspondences.
- We propose an adaptive matching approach that integrates features and matching rules. A matching process is iteratively extended, rewritten and executed. The automatically computed process can also be edited by the user.
- We evaluate or approach on a broad set of mapping problems from different domains and show its robustness.

The remainder of this paper is organized as follows: Section 2 gives a short introduction into some important preliminaries on schema matching. In Section 3 our self configuring matching approach is introduced consisting of features, matching rules and adaptive process construction. After that, Section 4 describes our self configuring schema matching system and our library of features and rules. We evaluate our approach in Section 5 and review related work on adaptive schema matching approaches in Section 6. Finally, we draw conclusions and give an outlook in Section 7.

## 2. PRELIMINARIES

Before getting into the details of features, matching rules and adaptive matching process execution we first need to give some definitions of the foundations of schema matching starting with our view of a schema.

A **schema** consists of a number of schema elements. Each element carries a name, a data type, and optionally a description (called annotation) as well as instances. The kind of schema is not restricted and can refer to any metadata structure such as XML schema trees, ontologies, database schemas or meta-models. The goal of a schema matching system is to compute a mapping suggestion between a source schema S and a target schema T. For computing the mapping, most matching systems use several matchers as well as other operators for aggregation and selection.

The **matcher** operator computes a similarity value for each pair of schema elements from the source schema S and the target schema T and constructs a **similarity matrix** of size $|S|*|T|$ as output. An entry in the similarity matrix is a value between 0 and 1 that represents the similarity between two elements with 0 representing low and 1 representing high similarity.

Most currently promoted matching systems use a combination of different matching techniques for improving the quality of matching results. For that purpose an **aggregation** operator is used. It combines results of multiple similarity matrices computed by different matchers to a single aggregated similarity matrix.

Finally, a **selection** operator extracts the most probable element pairs from a similarity matrix and sets all other values to zero. A number of selection strategies were proposed in literature [15]. The most important ones are Threshold, Delta and MaxN. Threshold simply filters all entries higher than a given threshold. The MaxN-Strategy returns the N highest entries in each row or column of a matrix; Max1 only considers the element with the highest similarity value as a match candidate. Delta extends MaxN by a delta environment around the N highest values of a row or column in a similarity matrix. All entries within this delta environment are added to the MaxN selection result. With the mentioned selection approaches, an element can be part of several correspondences as useful for 1:n, n:1 or n:m mappings. From the finally selected matrix a **mapping** between a source schema S and a target schema T can be constructed. A mapping consists of a set of correspondences (s, t, sim) referring to a source- and a target element as well as a similarity value.

Matching systems not only differ in the implementation of these basic operators but also by the order in which these operators can be executed. In this paper we adopt the notion of matching processes similar to eTuner [23]. A **matching process** (or matching strategy) is represented by a directed acyclic graph describing the execution order of operators such as match, aggregation or selection. It contains all steps necessary to come from two input schemas to a final mapping. Operators in the graph get one or more similarity matrices as input and return a similarity matrix as output. The topology of a matching process can be very different. Simple topologies that are commonly used are parallel, sequential and iterative execution of matchers as visualized in Figure 1.

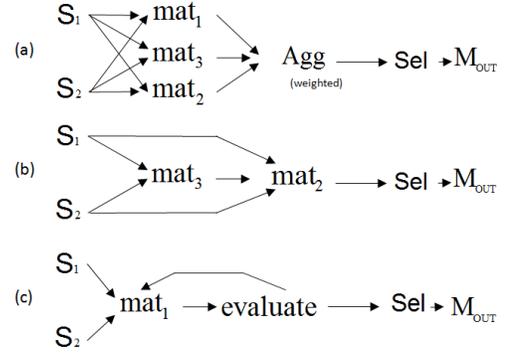

**Figure 1: Topologies: (a) parallel, (b) sequential, (c) iterative**

However, these basic topologies may also be combined in more sophisticated match strategies.

In general, tuning a matching system involves defining the underlying matching process structure, selecting the appropriate operators and parameterizing individual operators. This leads to a huge space of possible configurations of a matching system. Currently the tuning is done manually and requires a lot of testing and matching experience. In this paper, we introduce a system that is able to automatically choose promising matchers, aggregation and selection operations and their parameters. Also the structure of the matching process can be automatically extended. The user of our system does not need to tune before solving a new matching problem.

## 3. ADAPTIVE MATCHING APPROACH

In order to achieve full automation in the construction and configuration we took inspiration from how a matching expert interactively develops and executes a matching process. After analyzing the source and target schema, she selects appropriate matchers and constructs an initial matching process. The process is executed and the result is inspected. Depending on that result, certain parts or parameters of the matching process can be changed and extended manually.

Our approach performs similarly but in an automatic way. In order to automate the analysis step, we rely on so-called features. These are computed from the input schemas but also from the intermediate mapping results. Features try to give some indication about schema properties or the quality of a mapping. Based on the computed features so-called matching rules are defined that represent expert knowledge about a relation between features and operators or process patterns. Finally, an adaptive process execution system selects and applies rules and incrementally executes the constructed process. In the following subsections features, matching rules and the adaptive process construction are introduced.

### 3.1 Features

In general a **feature** takes one or several schemas or similarity matrices as input and computes a value between 0 and 1 as output. We distinguish between schema features, and matrix features. The notion of matrix features is newly introduced by this paper.

**Schema features** try to describe properties of schemas and can be computed in a preprocessing step before actually executing a matching process. In simple cases they reflect the schema size or the relative frequency of schema element properties such as the availability of element descriptions or data type information. More complex features rely on value distributions of schema elements or structural properties. For instance, the average length of paths in a schema tree gives some indication on when to use a path matcher evaluating the name similarity of elements and their predecessors. Some schema features evaluate the degree of similarity between both input schemas. For example the structural and linguistic schema similarity can be used to decide about the appropriateness of applying a structure-based or name-based matcher [14].

We additionally analyze intermediate similarity matrices after executing operators of the matching process to derive **matrix features**. They are used to evaluate the quality or similarity value distributions of similarity matrices. For instance a so called Noise feature computes the number of low valued entries in a similarity matrix in relation to the top-1 values in each row and column. The resulting feature can be used to evaluate the quality of a matrix and thus the operator that has generated it. Some matrix features take more than two matrices as input. They often describe the degree of commonalities and differences between multiple result matrices. For instance a feature could measure the overlap of top-1 values of different similarity matrices. If the overlap is higher, more confidence could be put in the different matrices.

In general, schema as well as matrix features formalize the results of a manual analysis step that a matching expert would generate before or while constructing a matching process, e.g. to select and add matchers.

## 3.2 Matching Rules

We use schema and matrix features within so-called matching rules. A matching rule captures a design decision a matching process expert would take in specific situations to increase the quality of a process for a given mapping problem.

A **matching rule** consists of the following parts:

- A **pattern** that describes a part of a process graph where the rule can be applied to. The pattern can be empty, in particular within rules that start the process construction.
- An **action** that applies a defined change to instances of the found pattern. This includes additions and changes of one or many (additional) operators to a process.
- A **relevance** function that computes the relevance of the respective rule for the current matching process and match task. It is based on computed schema and matrix features on the input schemas and already computed similarity matrices and computes a relevance value between 0 and 1. The relevance is used to decide whether a rule is executed.
- An optional **check** function that is used after a rule was applied to a process. It rates the quality of the changes that were introduced by the action. It also relies on matrix features to compute a value between 0 and 1.

To better explain the parts of a matching rule we introduce a simple example rule for reducing noise of matcher operators. The rationale of the rule is that a number of low-valued similarity entries in a result matrix often negatively influence a later aggregation. Reducing the noise often increases result quality. Figure 2 visualizes the pattern and action of the rule. On the left, our notation for rules is shown with the pattern above and the change of the action below the bar. Additionally the features for computing the relevance and the check are listed. In the example the pattern describes a process part that consists of a matcher operator (mat) that has some arbitrary following and preceding operator. When the rule is applied to a process part, all found instances of the pattern are collected.

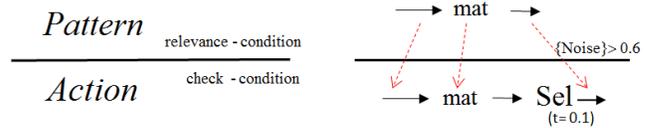

**Figure 2: Example matching rule**

For each matcher operator the result matrix is analyzed using the Noise feature from above. If the computed relevance is higher than the given threshold 0.6 a selection operator is inserted after the matcher operator. The selection operator gets a very low threshold t=0.1 to not prune out valuable similarity values. The new selection operator is executed. In this rule no check function is needed. However, in other cases the after-check can be used to ensure that the rule application did improve the matching quality, measured by features. If the check is negative the changes of the action are reverted.

We identified different types of rules that are shortly described in the following. As will be discussed below and visualized in Figure 3 the different kinds of rules are applied in a certain order to control the degree of process adaptations.

**Starting rules** can be applied to an empty process when no intermediate matrix was computed yet. In our system, starting rules mostly add basic matchers to the matching process that only take individual node attributes into account when computing similarities. Each application of a starting rule creates so called dangling nodes that are possible end-points of the process and do not have following nodes. At the dangling nodes the process is further extended in the following steps.

**Aggregation rules** add aggregation operators to a process and combine a number of dangling operators from a matching process. Dangling operators could have been created by starting rules. A multitude of aggregation operators exist such as AVERAGE [3], OWA[13], HARMONY[16] or MIN/MAX [3], each with advantages and disadvantages.

**Rewrite rules** take a non-empty matching process MP as input and rewrite the process to a new matching process MP'. Rewrite rules change the structure of a given process without changing the input and dangling output nodes. For instance the order of operators could be changed or additional operations can be added in between others. The noise reduction rule is an example for a rewrite rule.

**Refine rules** add operators to dangling operators in order to increase result quality. Some rules rather strive for precision whereas others try to increase recall. For instance, some refine rules add structural matchers to a matching process to propagate found node matches and identify additional matches that can be derived from structural similarities.

Before finalizing a matching process, **selection rules** can be applied. They are used to add a selection operator to the last dangling node of the current matching process. As with the aggregation operator, a number of selection strategies were proposed in literature [15].

## 3.3 Adaptive process construction

Obviously different rule classes depend on each other since some rules add operators and dangling nodes and others combine the output of several operators as done by the aggregation rule. In order to restrict complexity we perform the application of rules only within a fixed number of stages as shown in the left part of Figure 3. This reduces the structural diversity an adaptively created process can have but simplifies the rule selection process significantly. If all rules would be able to compete in all stages of the process side-effects of rule application could not be controlled and termination could not be ensured.

The process starts with importing the input schema and analyzing them to compute the schema features. An empty matching process is created. In the next stage starting rules can be selected and applied. Starting rules mostly add element-based matchers.

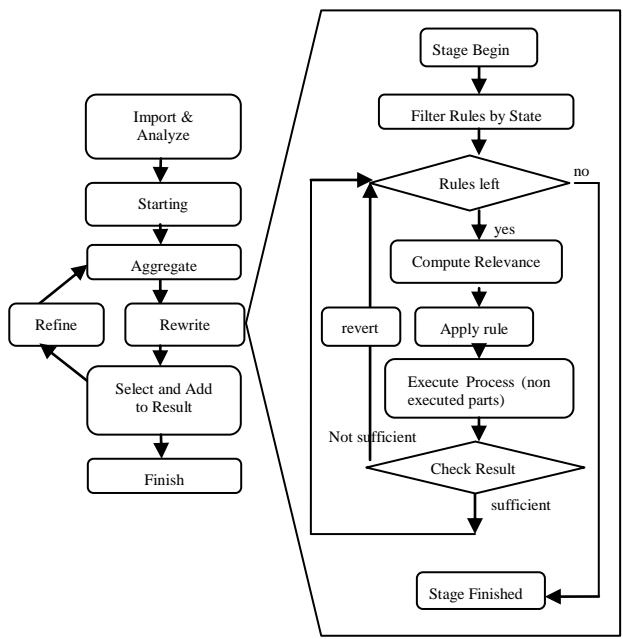

**Figure 3: Stages and rule selection**

However, also more complex starting rules can be defined that already construct an advanced matching process structure, e.g. to enforce the sequential execution of several matchers. In the next stage the dangling nodes from applying the starting rules need to be combined and the result matrices are aggregated. After the aggregation, rewrite rules can be applied. If no relevant rewrite rules can be found, a selection rule is applied. Based on the selection result the process can be finished or refine rules can be applied iteratively.

Within each stage of the process a predefined rule selection process is started (see right state chart from Figure 3). The selection process begins with filtering all rules that can be executed within the current stage. This set of rules is only created once within a stage. If the remaining set of rules is empty the stage can be finished directly. If there are rules left to be applied, their relevance is computed for each rule using the rules relevance functions. The most relevant rule is selected and applied (3). Applying a rule implies changes to the current matching process. After that the current process is executed. However no operator is executed twice and only new or changed parts are executed. The matrix result of executing the most recently applied rule is evaluated using the rules check function. It often happens that rules are rated as relevant due to the existence of certain attributes in the source and target schema. However, after executing the matchers that were added by the rule the matrix result quality is sometimes very low, indicating that the most recent rule should be ignored. For that purpose the most recent rule effect is reverted. After executing a rule it is removed from the remaining list of rules. Again the most relevant rule is selected and the process runs on until the rule set is empty.

## 4. ADAPTIVE MATCHING SYSTEM

We have developed a matching system that supports the execution of matching processes and implements the proposed adaptive matching approach (see Figure 4). To solve a match problem, the matching system gets two schemas as input and returns a mapping as output. Ideally no further parameterization input should be needed when running the system. All necessary parameters should be defined automatically.

The system consists of a registry that contains a number of feature analyzers, matching rules as well as an operator library that contains all necessary operators in particular the matchers, aggregation or selection operators.

The core component of the system is the adaptive process construction that basically implements the proposed staged rule application approach. In a preprocessing step all schema features of the input schemas are computed and cached to avoid double computation. After every change of the process the matching process execution is called to execute the new operators. This creates new intermediate similarity matrices that can be analyzed by subsequent matrix features. Currently the adaptive execution always starts with an empty process. In the future we plan to also support the adaptation of already existing matching processes.

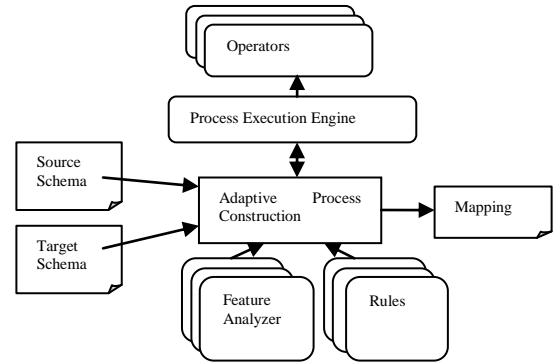

**Figure 4: Adaptive Matching System**

In the following, each of the components is described. Moreover, the most important feature analyzers and matching rules are introduced in detail.

### 4.1 Feature Analyzer Library

The feature analyzer library offers a set of analyzers to compute different schema and matrix features. Schema features are mostly used for computing the relevance of a rule. Matrix features can be used in both the relevance computation as well as the check function after a rule has been applied and the modified process has been executed. An important property of feature analyzers is that their computational complexity should be low to reduce their impact on execution efficiency. The library currently contains more than 20 feature analyzers as shown in Table 1 together with the required input. Some of these are very simple existence features that check availability of certain properties for matching like datatypes or annotations. However, also more advanced analyzers are included that analyze the distribution of matches between two schemas or compute a structural similarity. The library of feature analyzers can easily be extended.

**Table 1: Features in the Library**

| Schema Features | Input | Matrix Features | Input |
|---|---|---|---|
| *Existence | S ∨ T | CrossMatches | M, S ∧ T |
| NodeTokenRatio | S ∨ T | MatchDistribution | M, S ∧ T |
| NameMeaningfulnes | S ∨ T | Harmony | M |
| PathVariance | S ∨ T | MultiMappings | M |
| RepeatingElements | S ∨ T | Monogamy | M |
| SchemaDepth | S ∨ T | Selectivity | M |
|  |  | Noise | M |
| SimilarLanguage | S ∧ T | Complementarity | M* |
| *TokenOverlap | S ∧ T | Unanimity | M* |
| NameSimilarity | S ∧ T |  |  |
| StructuralSimilarity | S ∧ T |  |  |
| StructuralContain | S ∧ T |  |  |

In the following for each class of analyzers the most important ones are introduced.

**Schema Features**

*{Name, Datatype, Annotation, Instance}-Existence* specifies the percentage of elements that carry a {Name, Datatype, Annotation, Instance}.

*NameMeaningfulness* was originally proposed for Rimom [14] to assess which percentage of schema element names is meaningful. This feature is implemented using a dictionary such as WordNet for looking up element names or their components. In the OAEI Benchmarks, some schemas were artificially changed by scrambling labels. Based on that feature, Rimom was able to entirely skip any name matching in such cases.

*NodeTokenRatio* analyzes the names of schema elements. It often occurs that schema designers only use a small set of terms and concatenate them to name schema elements. This easily creates ambiguity. With the help of this feature, an appropriate name matcher such as TF/IDF can be chosen that tries to include the relative importance of terms into the computation of name similarities.

*RepeatingElements* measures how often element names and their content are repeated within a schema. In particular in XSD schemas, types are often reused which creates ambiguity and high values in the repeating elements feature.

*{Name, Annotation}-TokenSetOverlap* tries to determine how similar the set of names or annotations from the two input schemas S and T are:

$$f_{tokenSetOverlap} = \frac{|ts(S) \cap ts(T)|}{|ts(S) \cup ts(T)|}$$

with $ts(S)$ being the set of all tokens from all element names or annotations of a schema S. High overlap values indicate that a *{Name, Annotation}*-based matcher could provide good matching results.

*StructuralSimilarity* was already proposed in Rimom [14] and was slightly adapted in UFOme [20]. It is a lightweight measure to compute how similar the structural shapes of two schemas are. A high structural similarity is an indicator to increase the relevance of structure-based matchers.

All proposed schema features can be computed before process execution. Additionally we need matrix features that are computed while executing the process.

**Matrix Features**

*Selectivity* tries to evaluate the confidence of a result matrix that was computed by a matcher or sub process. It computes the distance of the top-1 entry in a row or column to the next highest entry in the same row and column. The rationale is that a high distance of the best candidate match to the next possible matches implies that the candidate match is certain. A low distance on the other hand shows more uncertainty. For a vector V sorted in descending order (so that $V_0$ is the similarity of the best, and $V_1$ of the next best candidate) we compute the selectivity of the vector as:

$$selectivity(V) = \begin{cases} 0 & if\ V_0 = 0, \\ V_0 - V_1 & else. \end{cases}$$

For a similarity matrix M with n×m entries the selectivity value $f_{selectivity}$ can be computed as follows:

$$f_{selectivity} = \frac{\sum_{i=1}^{n} selectivity(M_{i,*}) + \sum_{j=1}^{m} selectivity(M_{*,j})}{2 * |\{m_{i,j} \mid m_{i,j} > 0\}|}$$

All selectivities of rows and columns are summed up and divided by the number of candidate entries in the matrix. If the selectivity is very low, the likelihood that after a selection many 1:n, n:1 or n:m matches (so called multi matches) will result is very high. For example, a high selectivity indicates to use a Max1 selection when a selection strategy needs to be defined.

*CrossMatches* computes how structurally consistent a computed mapping is, i.e. how structurally close the matching target elements of structurally close source elements are. A low structural consistency is an indicator for low precision mappings. In order to increase structural consistency special constraint based selection approaches as proposed in ASMOV [12] could be used.

*MultiMatches* represents the ratio of multi matches to the number of 1:1 matches on an already selected similarity matrix. This feature is later used to compute the relevance of rules that reduce the multi matches in order to increase quality.

*Monogamy* computes how close all found mapping pairs in a matrix are to a monogamic relationship. The feature was inspired by the *Harmony* value that was proposed in [16] for automatically deriving aggregation weights. Harmony counts the number of entries in a similarity matrix that are maximal entries both on its rows and columns. Since many matchers compute matrices with many multimatches the harmony value is often very low even though the final mapping quality might be good. In order to cope with multimatches the monogamy value can be used. In a monogamic relationship each partner of a match pair should only be involved in this and possibly no other match relationship. The more each partner is involved with other partners, the lower the computed monogamy value is. Therefore the first mapping in

Figure 5(left) has a low monogamy value, whereas monogamy for the second mapping is higher.

Monogamy is closely related to the stable marriage property [10]. However, it is not restricted to 1:1 mappings but allows to measure how close existing n:m relationships are to a 1:1 relationship. In the evaluation we can show that this value is a very robust indicator of a good quality mapping result, even though the actual F-Measure cannot be computed without a gold standard.

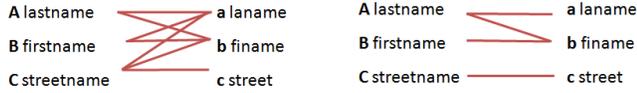

**Figure 5: Monogamy: distance to a stable relationship**

For mapping problems where the correct mapping is not 1:1, the monogamy should not be applied. However by using the multimatches and repeating elements features we can predict the most probable kind of mapping. To further explain the monogamy measure, Figure 6 illustrates with an example how the monogamy feature is computed.

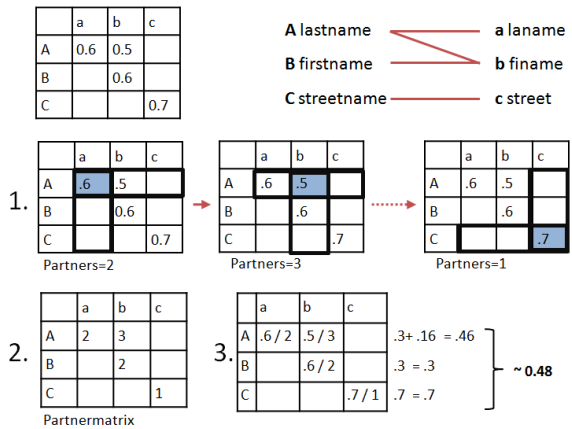

**Figure 6: Monogamy computation**

At the top the initial matrix and the corresponding mapping is visualized. The algorithm iterates through all entries of the matrix and counts the number of partners in the row and column (1). The partner number for each entry is put into a separate partner matrix (2). Finally each similarity value is divided by its partner count which weights entries by the number of partners (3). The weighted similarities are then summed up by row and finally averaged to the resulting monogamy value.

Certainly also other matrix features can be computed such as the average of all similarity values in a matrix or the DICE-Value [3]. However these features can only be used in combination with others since the information that can be derived from these values is low.

### 4.2 Rule registry

Based on the presented features a number of rules were defined that are put into the matching rule registry. The rule registry can be extended by additional rules. Currently many rules are simple and add a specific matcher if the associated relevance value is high. Before adding complex rules we need to make use of the basic decisions a matching process expert would make. However, we have also started to add more complex rules that add more than one operator to a matching process. Table 2 gives an overview to several rules of our library ordered by their rule type. Start rules are typically used to add simple element-level matchers based on name, type, annotation and instance similarity while structural matchers are added by refine rules.

**Table 2: Matching Rules in the Library**

| Matching Rule | Type | Matching Rule | Type |
| --- | --- | --- | --- |
| AddInstanceMatcher | Start | AddSelectDelta | Select |
| AddWeightedNameMatcher | Start | AddSelectMaxN | Select |
| AddTokenNameMatcher | Start | AddAverageAgg | Agg |
| AddDataTypeMatcher | Start | AddPathMatcher | Refine |
| AddAnnotationMatcher | Start | AddSiblingMatcher | Refine |
| AddNoiseReducingSelect | Rewrite | AddChildrenMatcher | Refine |
| SequentialRewrite | Rewrite | AddParentMatcher | Refine |
| AddBlockingMatcher | Rewrite | AddStatisticsMatcher | Rewrite |

In the following we briefly describe selected rules.

*AddWeightedNameMatcher* rule adds a term weighting feature to the name matcher. This is necessary since schema designers often use a restricted set of terms for naming schema elements. The weighting approach is able to reduce the importance of terms based on their occurrence counts similar to TFIDF in information retrieval. However weighting should only be used under special circumstances. These are measured by the relevance function that relies on the NameExistence, NodeTokenRatio and RepeatingElements features. Certainly if no names are set the NameExistance computes low values, leading to a low relevance value. If names can be found the NodeTokenRatio should be smaller than 0.8 and the RepeatingElements bigger than 0.8. The rationale behind that is that repeating elements could also lead to a lower NodeTokenRatio. However in that case the weighting could decrease matching quality. After executing the changed process the rules check function is used. Here we rely on the monogamy value. If the value is lower than a small threshold of 0.1, the rule is rolled back, by removing the added operator.

*AddPathMatcher* introduces a path matcher taking the currently computed similarity matrix as input for computing the path similarity. The input is often referred to as constituent matrix. The relevance function of the PathMatcher rule relies on the PathVariance, SchemaDepth, Selectivity as well as the MultiMatches feature. If the SchemaDepth value is very low in one of the input schemas a flat schema structure can be assumed. That leads to a low relevance of the PathMatcher rule. If the PathVariance is high, the MultiMatches feature has a high value and the Selectivity is low then also the relevance for the path matcher gets a high value. The means that the path matcher is particularly useful to increase the selectivity of a similarity matrix or reduce the number of multi matches. The check function computes the monogamy value and additionally checks whether the Selectivity value increased and the MultiMatches value decreased. The rule was inspired by [4]. In their work a fixed filtered context process was described that takes multi mappings of node-based matcher and applies a name-path matcher to them, to resolve the found multi mappings to 1:1 correspondences.

*AddSelectDelta* is a rule that can be applied after each refine iteration. Its relevance is computed from the MultiMatches, MatchDistribution and SchemaSizeRatio feature. If there are multi matches involved and the difference of schema sizes is high then the probability that multi matches should be part of the final result is higher. Additionally we evaluate if the distribution of matches across the bigger schema is equal or if multimatches are structurally close. For instance a source element could match to both a target element and its parent. Certainly in that case only one of the two matches should be taken for the final result and a maxN=1 selection should be added instead. The delta selection allows to include additional matches to the maxN=1 selection. However defining the delta value is complex and different from use case to use case. Hence for each use case, we test different delta values and compute the monogamy value for the possible selection result. The delta value producing the highest monogamy value is chosen for the selection operator. In the evaluation we show that this adaptive computation of the delta value increases result quality.

For space reasons, not all matching rules can be described in detail. However, in general, the rules relevance functions rely on features that best project the possible result quality of the added operator. In the check function we make heavy use of the monogamy value. In particular for the node based matchers it helped to drop irrelevant matchers.

### 4.3 Adaptive Execution Example
For better understanding we explain the adaptive construction when matching two anatomy taxonomies that will also be used in the evaluations. In Figure 7 different states of the adaptive construction are visualized.

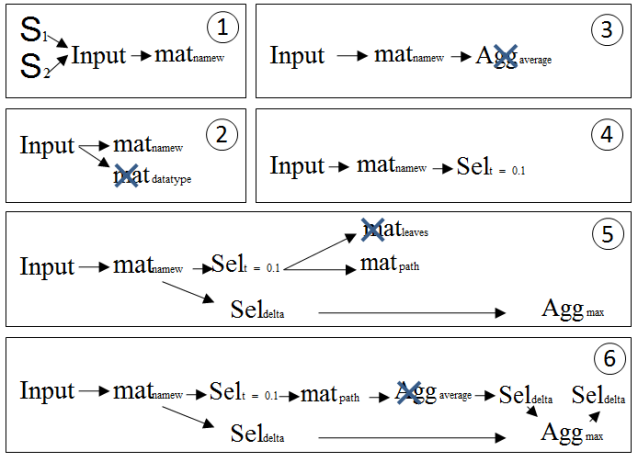

**Figure 7: Example adaptive construction**

Our system starts with constructing a new process with an input node that imports the source and target schemas and computes the schema features. In (1) the *AddWeightedNameMatcher* starting rule was applied due to a high number of repeating tokens in element names. In (2) the *AddDataTypeMatcher* rule was relevant and executed. However it was then removed due to a low quality computed by the check function. After all relevant starting rules were executed an aggregation rule can be executed (3) that combines dangling nodes. In the example, only one dangling node exists so that the Aggregation operator can be removed. In (4) the *AddNoiseReducingSelect* rewrite rule was applied and a selection operator was added after the matcher operator. In (5) the *AddSelectDelta* rule was executed and a final Aggregation operator with a maximum aggregation strategy was added.

Additionally two refine rules were executed that add a path and a leaves matcher. The *AddPathMatcher* rule is selected since the result matrix of the name matching still contains some multi matches as measured by the MultiMatches feature. In (6) the *AddSelectDelta* rule was applied adding its result to the existing Aggregation node.

## 5. EVALUATION
In the evaluation we want to investigate the effectiveness and robustness of our approach. For that reason we compare our adaptive schema matching system to currently known alternative approaches for diverse problems.

### 5.1 Test Data
We consider a wide range of schema mapping problems from different domains. To be comparable to existing work we also include the OAEI-Benchmark and the ModelCVS Benchmark in our evaluation. Table 3 lists the used datasets together with the number #M of considered mapping scenarios. The Purchase Order dataset and mappings were already used in the early evaluations from COMA [3] and they are publicly available. We also use them for computing a default configuration to compare against.

**Table 3: Test datasets**

| Testsets | #M | Schema Types |
|---|---|---|
| Purchase Order | 10 | Small XSD schema mapping problems |
| Enterprise Services | 52 | Small and large XSD, IDOC mappings |
| OAEI Benchmark | 110 | Synthetic ontology mapping problems |
| Anatomy | 1 | Large-sized taxonomy mapping problem |
| ModelCVS | 10 | Hard metamodel matching problems |

Additionally we can compare us indirectly to the published COMA results for that data set. The *Enterprise services* dataset contains a big set of mappings between service interfaces from an SAP Enterprise Services Repository. The set contains a big diversity of problem sizes and complexities. In particular the mappings involving SAP Intermediate Document Formats (IDOC) are challenging since the names are often cryptic. The *OAEI Benchmark* consists of 110 smaller synthetically generated ontology mappings. These mapping problems are often artificially created. Also the *Anatomy* mapping is provided by the OAEI. It is a relatively large real-world mapping scenario with very high schema similarity. About 60 percent of the correspondences are trivial due to a very high name similarity. Finally we use the *ModelCVS* dataset from [9] that provides metamodel matching problems that differ strongly in the way elements are named and structured.

### 5.2 Setup
We precomputed a best configuration for the Purchase Order dataset similar to the way the default strategy proposed in [3] was computed. For that purpose we generated all possible matcher combinations and created a parallel matching process. We tested different selection strategies and delta parameters to find an optimal matcher combination and selection strategy. Due to the huge space of parameter settings and combinations this process took several days to finish. We call this computed strategy our DEFAULT configuration. It consists of the matchers WeightedName, Path, Children, Leaf, Sibling and Datatype. The best selection strategy found was parameterized with delta=0.021 and a threshold=0.5.

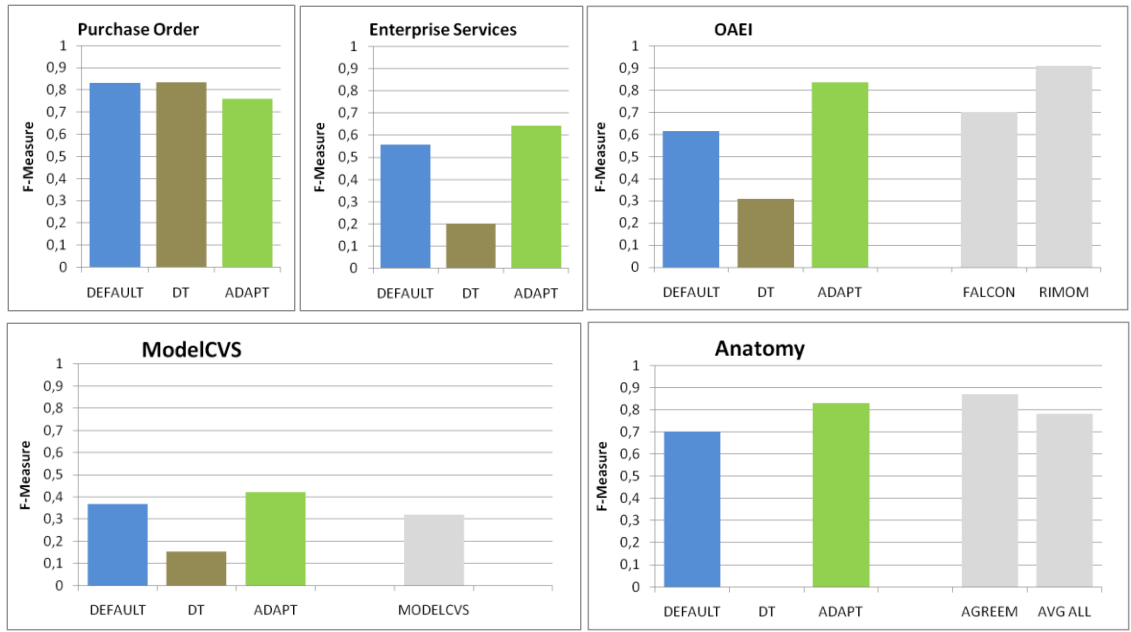

**Figure 8: Evaluation results for adaptive execution**

Secondly we implemented an alternative, learning-based approach called meta level learning that was proposed in [5]. This approach is a valid comparison since it includes schema features in a learning process of a decision tree to increase adaptivity. The learning also takes the Purchase Order dataset as gold mappings. We include all schema features in the learning process. The learning was implemented using the weka[1] library. The computed decision trees are then translated into matching processes without loss of information. In order to reduce possible overfitting we restricted the size of the decision tree. We then executed the default configuration, the decision tree and our new adaptive matching system with all provided mapping problems.

## 5.3 Overall Results

The results of our evaluations are shown in Figure 8. For each dataset we compared the average achieved F-Measure of the individual approaches DEFAULT, DT and ADAPT. For the Purchase Order data set the adaptive approach (ADAPT) is only closely behind the DEFAULT configuration and the computed decision tree (DT). The slightly better outcome for DEFAULT is not surprising since it was computed on this data set by testing all possible parameter configurations. Also DT was learned on that data set. However we can already see that it is hard even for a learned process to be better than the DEFAULT configuration since the diversity of matches is too big to derive a representative rule in a decision tree. If the size of the tree would not be restricted the results get slightly better. However, then we can assume that the strategy will overfit which was already experienced by the authors of [5].

[1] http://www.cs.waikato.ac.nz/~ml/weka/index.html

In the Enterprise Services test cases our adaptive strategy achieves about 10% better results than the DEFAULT strategy. This is due to the adaptive selection of the appropriate matchers and the automatic definition of the selection delta. A number of schemas in the Enterprise Services dataset contain annotations. In order to be comparable we simply added an annotation matcher to the DEFAULT matcher configuration. However the increase in F-Measure was only about 2 percent. Surprisingly the learned decision tree DT failed to return reasonably good results. In the Purchase Order test cases many correspondences are trivial in that average similarity values of matches are close to 1. This is also detected by the decision tree so that a lot of correct matches with lower similarity are pruned early in the tree. Obviously the DEFAULT strategy is much more robust regarding that problem.

In the OAEI test cases we achieve a reasonably high F-Measure. Here we are nearly 14 percent better than the DEFAULT strategy. DT again fails to return good results. Again we added an Instance and an Annotation matcher to the DEFAULT strategy in order to be comparable. But almost no quality increase was measured since the strict selection thresholds filtered out possible additional correspondences. In order to be able to rate the quality of the adaptive process we added the results from Falcon and Rimom from the recent OAEI contest. Rimom achieved an F-Measure that is 7% better while Falcon performed worse than our strategy. We also constructed a manually defined matching process that achieves F-measure values around 92% . We did not include this in the result comparison since the goal of the adaptive approach is automatic adaptivity and robustness for diverse problems, not achieving the maximal possible F-Measure for a single data set.

For the ModelCVS case the differences to the DEFAULT strategy are smaller. However this is also due to the difficulty of the mapping problem. In some cases the F-Measure is lower than 0.2 so that rules based on a projected a result quality are skipped although they might have worked well. However we still win more than 10 percent to the best achieved matching results achieved by the ModelCVS group [9].

Finally in the Anatomy case we again win around 13 percent in comparison to a default strategy. When comparing to the maximal possible values achieved by AgreementMaker (AGREEM) [2] that are around 87 percent we are quite close to an optimum. Also we are above the average F-Measure achieved by systems participating in the anatomy track of the recent OAEI campaign. Unfortunately we were not able to run the decision tree due to the size of the matching problem. Currently we run into memory issues due to the high number of match nodes in the tree.

In summary we could show that the adaptive approach was able to be better than the considered competitor approaches for almost all data sets. If strategies are tuned for a specific dataset like DEFAULT or DT one can achieve better results individually. However, due to the size of existing data sets a process as proposed for computing the default computation is not feasible for the other mapping problems.

### 5.4 Adaptive Selection

After having shown the strength of the adaptive matching approach we now want to show the value of the newly introduced monogamy value. For that purpose we ran a matching process on two scenarios and iteratively changed the threshold of a threshold-based selection strategy at the end of the process. With each threshold we computed the monogamy of the result matrix. The result is shown in Figure 10.

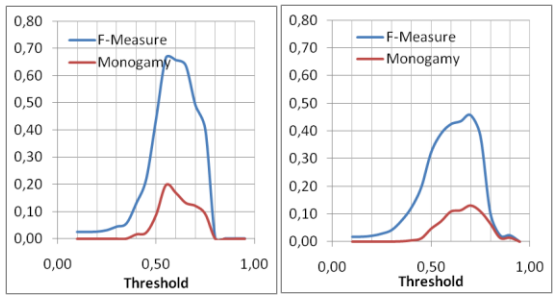

**Figure 9: Using Monogamy to define threshold values**

What can be seen is that there is a very strong correlation between the F-Measure of the strategy and the computed monogamy value. The two examples are taken from the Purchase Order dataset. However, if only few n:m correspondences are included the monogamy can also help to derive a delta value for delta-selection operator. Instead of a threshold we changed the delta value from 0 to 0.2 and measured the monogamy. Again, choosing the maximum monogamy value gives some indication for a probably good F-Measure. Note that the monogamy works well for 1:1 matches that may have a few 1:n correspondences. For all other cases the monogamy should not be used for finding a delta value. In these cases, a static value would be preferable. Since we already have a number of features available we can predict when to use the adaptive delta selection.

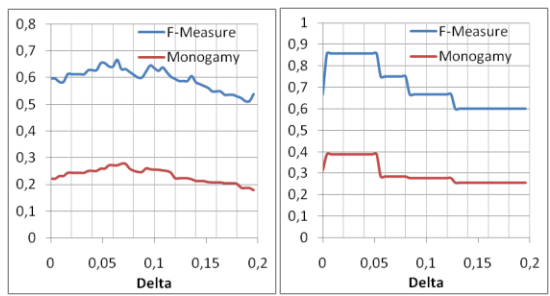

**Figure 10: Using Monogamy to define delta values**

### 6. RELATED WORK

Different approaches were proposed in the past to make schema matching systems more adaptive. First attempts tried to increase adaptivity on the operator level. The AVERAGE aggregation operator and the DELTA selection strategy from [3] were already quite robust without the need to analyze schemas and mappings. This was also reproduced in our evaluation with the DEFAULT strategy. A first approach that also relies on the analysis of the mapping was the harmony based aggregation as described above [16]. However, the usefulness of this approach seems restricted to the OAEI Benchmark and we could not reproduce the effects on other data sets. However the harmony value served us as an input to our newly developed monogamy feature.

Another promising attempt to adapt matching systems to the problem was introduced by eTuner [23]. This system is able to construct a synthetic gold standard by iteratively perturbing one of the input schemas. Since the perturbation rules are known, a mapping between the original source and the perturbed schema is known. This synthetic gold standard is then used to tune a given matching process. However, their proposed approach seems not feasible in practice. First they ignore the target schema and perturb only the source. The perturbation rules are static so that the generated gold standard does not differ much for different mapping problems. The synthetic gold standard could differ a lot from the original mapping problem, so that the system will be tuned wrongly and return weak results.

First approaches were proposed that use a rule based selection of matchers, selection and combination operators. However, these attempts are very restrictive. Either they fully rely on a questionnaire [17] that involves a lot of manual user input or they use hard-coded rules as Rimom [14] or Falcon [11] does. These two systems were among the first to exploit a preprocessing and analysis of the input schemas to be matched. Based on structural similarity or name similarity, an edit-distance- or structural matcher is included in the matching process. These selection criteria are fixed into code while we support a modular adaptation approach based on separately managed matching rules.

Meta level learning [6] was the first to recognize the need to have more schema features for creating adaptive processes. However the problem is that in practice, often no or no suitable gold mappings are available for learning. Additionally the mapping problems a system is faced with differ a lot so that learned models often are not able to return results with a good quality. The authors acknowledge that their learning approach easily overfits with the learning base. Also with increasing sizes of decision trees the performance drops significantly. Other learning techniques like YAM [5] or [7] might not suffer that strongly from overfitting, but they do not consider schema features. YAM proposes to apply different classifiers for different mapping use cases. However, the user is asked to select the appropriate classifier or to use a default classifier learned over a huge mapping knowledge base. Our system does not propose a default strategy. Also the user is not involved in the selection of appropriate strategies.

Expressing knowledge as rules was also proposed by the UFO-ME [20] strategy prediction module. However they mainly re-implemented the rules and internal process defined in Rimom. Also they did not consider rule selections and relevance computation. Additionally their major goal was to provide the user with the predicted strategy for further editing. In contrast to that our approach does not involve the user in the configuration difficulties. Others introduced a rule-based rewrite approach for matching processes [19]. Their approach was restricted to performance optimization and ignored the quality aspect. However, the proposed rules could be integrated as rewrite rules to our rule registry. To our knowledge, our approach is the first to introduce matrix features and rules that rely on these features.

## 7. CONCLUSION&OUTLOOK

We proposed a new self-configuring and adaptive schema matching system that is able to return good mapping results for very different schema mapping problems. Our system relies on a number of schema and matrix features that are computed from the input schemas as well as from intermediate results of a matching process. These features are used in matching rules to select matchers, aggregation and selection operators and to adapt matching processes. An adaptive matching process selects and executes rules based on their computed relevance. This automatically creates a complex matching process that is adapted to the problem at hand. We also proposed the monogamy feature that provides an indication about the quality of a mapping without requiring a gold standard.

In our evaluations we could show the strength of our approach. With our system we were able to compete with other manually tuned matching systems even though we intended to achieve good matching results across different mapping problems.

The proposed approach is in its initial version so that there is still room for improvement. First, the current set of rules can be extended to support more complex adaptations. E.g. to split process control flow based on the type of elements to be matched. Each branch is then tuned differently. Secondly we began to integrate a learning solution into our system that allows learning the relevance functions of individual rules. This might be a promising combination of learning and adaptive matching since currently the relevance functions are defined manually when the rule is designed. Finally we want to further close the gap to manually tuned matching processes by identifying additional rules and features.

## 8. REFERENCES

[1] Bellahsene, Z.; Bonifati, A.; Rahm, E. (eds.). 2011. Schema Matching and Mapping. Springer-Verlag.

[2] Cruz, I.F., Palandri, F., Stroe, C. 2009. AgreementMaker: efficient matching for large real-world schemas and ontologies. PVLDB Volume 2

[3] Do, H. H. and Rahm, E. 2002. COMA - A System for Flexible Combination of Matching Approaches. VLDB Proc.

[4] Do, H. H. and Rahm, E. 2007. Matching large schemas: Approaches and evaluation. Inf. Syst., 32(6).

[5] Duchateau, F., et. al. 2009. YAM: a schema matcher factory, CIKM.

[6] Eckert, K., Meilicke C., Stuckenschmidt, H. 2009. Improving ontology matching using meta-level learning. In ESWC 2009.

[7] Ehrig, M., Staab, S., Sure. Y. 2005. Bootstrapping Ontology Alignment Methods with APFEL. In WWW '05.

[8] Euzenat, J., et. al. 2010. Results of the Ontology Alignment Evaluation Initiative 2010. Workshop on Ontology Matching.

[9] Falleri, J.-R., et. al. 2008. Metamodel matching for automatic model transformation generation. MoDELS '08

[10] Gusfield, D., Irving, R. W. 1989. The stable marriage problem: structure and algorithms. MIT Press.

[11] Hu, W. and Qu, Y. 2008. Falcon-AO: A practical ontology matching system. Web Semant., 6(3).

[12] Jean-Mary, Y. R., Shironoshita, E. P., Kabuka, M. R. 2009. Ontology matching with semantic verification. Web Semantics Journal 7/3.

[13] Ji, Q., Haase, P., Qi, G. 2008. Combination of Similarity Measures in Ontology Matching by OWA Operator. IPMU'08

[14] Li, J. et al. 2009. RiMOM: A Dynamic Multistrategy Ontology Alignment Framework. IEEE Transactions on Knowledge and Data Engineering, 21(8).

[15] Meilicke, C., Stuckenschmidt, H. 2007. Analyzing Mapping Extraction Approaches. ISWC - Workshop on Ontology Matching.

[16] Ming M.; Yefei P.; Michael S., 2008. A Harmony Based Adaptive Ontology Mapping Approach. SWWS'08.

[17] Mochol, M., Jentzsch, A., 2008. Towards a rule-based matcher selection. EKAW '08.

[18] Noy, N. F. and Musen, M. A. 2003. The PROMPT Suite:Interactive Tools for Ontology Merging and Mapping. Int. J. Hum.-Comput. Stud.

[19] Peukert, E., Berthold, H., Rahm, E. 2010: Rewrite techniques for performance optimization of schema matching processes. EDBT.

[20] Pirrò, G.,Talia, D., 2010: UFOme: An ontology mapping system with strategy prediction capabilities. Data Knowl. Eng. 69(5)

[21] Rahm, E. and Bernstein, P. A. 2001. A survey of approaches to automatic schema matching. The VLDB Journal 10.

[22] Shvaiko, P and Euzenat J. 2005. A Survey of Schema-Based Matching Approaches. Journal on Data Semantics IV.

[23] Y. Lee et. al. 2007. eTuner: tuning schema matching software using synthetic scenarios. The VLDB Journal, 16(1).